\documentclass[final,5p,times,twocolumn]{elsarticle}

\usepackage{lineno,hyperref}
\modulolinenumbers[100]

\usepackage{hyperref}
\hypersetup{colorlinks=true,linkcolor=blue,filecolor=blue,urlcolor=blue,citecolor=blue}
\usepackage{orcidlink}
\usepackage{amssymb}
\usepackage{dcolumn}
\usepackage{graphicx}
\usepackage{tabularx}
\usepackage{bm}% bold math
\usepackage[T1]{fontenc}
\usepackage{amsmath}

\begin{document}
\begin{frontmatter}

\title{Large geometric polarization and magnetic behavior in the multiferroic quasi-2D SrNiF$_4$ fluoride}

%\tnotetext[mytitlenote]{acgarcia@uis.edu.co}

%% Group authors per affiliation:
\author{W. Ibarra-Hernández$^{1}$\,\orcidlink{0000-0002-5045-4575}} 
\ead{wilfredo.ibarra@correo.buap.mx}
\author{A. Bautista-Hernández$^{1}$\,\orcidlink{????}} 
\address{$^1$Facultad de Ingenier\'ia, Benem\'erita Universidad Aut\'onoma de Puebla, Apartado Postal J-39, 72570, Puebla  (Puebla) M\'exico.}

\author{A. C. Garcia-Castro$^{2}$\,\orcidlink{0000-0003-3379-4495}} 
\address{$^2$School of Physics, Universidad Industrial de Santander, Carrera 27 Calle 09, 680002, Bucaramanga (Santander)  Colombia.}
\ead{acgarcia@uis.edu.co}
%\fntext[myfootnote]{Since 1880.}

\date{\today}

\begin{abstract}
In the last decades, multifunctional single-crystals that show, for example, 
multiferroic and magnetoelectric responses have attracted 
considerable attention due to the potential applications and 
physical phenomena involved in the entanglement of the ferroic 
orders.
This paper investigates the structural, ferroelectric, and magnetic properties of the unexplored layered SrNiF$_4$ fluoride compound.
We show that, in terms of the vibrational phonon modes, this fluoride compound shows a tangible ferroelectric behavior with a spontaneous polarization as large as $P_s$ = 14.8 $\mu$C$\cdot$cm$^{-2}$ standing as one of the largest among its family. 
Besides, based on our findings such ferroelectric polarization presents a geometric origin that is strongly entangled with the octahedral rotations within the $\Gamma_2^-$ phonon mode and enhanced by the structure's $\Gamma_1^+$ mode. 
We also observed that the spontaneous polarization is coupled to the noncollinear $G$-type antiferromagnetic ordering in the structure allowing a switching of the weak antiferromagnetic component when the polarization is reversed.
\end{abstract}


\end{frontmatter}

%\linenumbers

%%%%%%%%%%%%%%%%%%%%%%
\section{Introduction:}
\label{sec1}
Research into ferroic orders in materials has revealed unforeseen capabilities, uncovering both physical and chemical phenomena. These discoveries have significantly enhanced the potential applications of advanced materials for future use \cite{Spaldin2019,Ramesh2020,doi:10.1098/rspa.2019.0542}.
Furthermore, the physical phenomenon related to two-dimensional (2D) ferroelectricity and 2D magnetism has been at the center of recent studies \cite{2D-ferroelectrics-2023, 2D-magnets-2019,doi:10.1021/acs.jpclett.9b01969}.
Here, bulk compounds showing a layered-type structure have been highlighted due to their possibility to exhibit tangible polar ferroelectric and magnetic orderings \cite{BWO-2019, PhysRevLett.121.117601, PhysRevB.90.064113, doi:10.1002/pssb.201570321, Benedek2015}. 
Nevertheless, the ensemble of polar oxide candidates within the layered stoichiometries, such as Ruddlesden-Popper, Aurivillius, and Dion-Jacobson, is quite limited, and therefore, similar materials that share this structural feature are in pursuit.

Among the ferroelectric, magnetic, and multiferroic compounds, several fluoride families and materials have stood as good prototypes for multifunctional applications for example including multiferroic and magnetoelectric behavior based on the ferroelastic-magnetic coupling \cite{PhysRevB.85.224430}, ferroelectric and magnetic behavior \cite{PhysRevLett.116.117202,Scott2011a}, geometrically-driven ferroelectricity \cite{PhysRevB.89.104107}, confined two-dimensional electron gases \cite{PhysRevB.102.235140}, and possibly, superconductivity at oxyfluoride interfaces \cite{Yang2014, Yang2015}.
Other properties also include applications in transparent ceramics with photonic response \cite{photonic-fluorides}, water-splitting processes \cite{Fan2018, doi:10.1021/acsaem.9b00449}, and catalysis phenomenon \cite{func-adv-fluorides, BABEL198577}.
All of the latter take advantage of the 2D-like response. 
As such, within this class of materials, layered structures with stoichiometric relationship $A_nB_nX_{3n+2}$ have been recently highlighted thanks to their structural behavior that allows the entanglement between the polar response and the octahedral rotations and tilting \cite{LICHTENBERG20011}. 
The latter is the case of the Ba$M$F$_4$ ($M$ = Ni, Mg, Cu, Zn, Co, Fe, and Mn) family where the spontaneous polarization and the antiferromagnetic ordering are structurally coupled \cite{PhysRevB.74.024102, PhysRevB.74.020401,doi:10.1021/acsami.5b10814}. Among these compounds, BaCuF$_4$ stands due to the exhibited strong Jahn-Teller distortion that favors the weak-ferromagnetic ordering and achieves an electrically-controlled magnetization effect close to room-temperature \cite{PhysRevB.84.075121,PhysRevLett.121.117601}. 
Therefore, this family of compounds represents a set of ideal platform candidates for the realization of single crystal magnetoelectric response.
With this in mind, further studies focused on the discovery of novel compounds showing 2D-like structures with possible enhanced properties are in the spotlight of investigation.

This paper presents the theoretical study, based on symmetry operations and density-functional theory calculations, on the ferroelectric, magnetic, and multiferroic response in the, experimentally synthesized \cite{SrNiF4-ref,BABEL198577} but unexplored, SrNiF$_4$ layered fluoride.
We have made use of the phonon modes analysis and we have explained the possible structural phase transition from the $Cmcm$ paraelectric to the $Cmc2_1$ polar phase, where the noncollinear antiferromagnetic ordering is explored and consolidates the multiferroic/magnetoelectric ordering.
In Section \ref{secII} we present and explain the computational details and theoretical approaches used for the development of this work. 
In Section \ref{secIII} we covered the discussion of our findings and the associated analysis.
Finally, in Section \ref{conclusions}, we portray our conclusions and general remarks.

%%%%%%%%%%%%%%%%%%%%%%
\section{Theoretical and computational details:}
\label{secII}
In our theoretical analysis, we stand in the density-functional theory (DFT) framework \cite{PhysRev.136.B864,PhysRev.140.A1133} within the implementation performed in the \textsc{vasp} code (version 5.4.4) \cite{Kresse1996,Kresse1999}. 
We used the projected-augmented waves, PAW \cite{Blochl1994}, approach to represent the valence and core electrons. 
As such, the electronic configurations considered in the pseudo-potentials for the calculations were Sr:4$s^2$4$p^6$5$s^2$ (version 07Sep2000), Ni:3$p^6$3$d^8$4$s^2$ (version 06Sep2000), and F:2$s^2$2$p^5$ (version 08Apr2002).
The exchange-correlation was represented within the generalized gradient approximation GGA-PBEsol parametrization \cite{Perdew2008} and the $d$-electrons in the 3$d$:Ni shell were corrected through the DFT$+U$, $U$ = 4.0 eV, approximation within the Liechtenstein formalism \cite{Liechtenstein1995}.  
The periodic solution of the crystal was represented by using Bloch states with a Monkhorst-Pack \cite{PhysRevB.13.5188} \emph{k}-point mesh of 12$\times$4$\times$8 and 600 eV energy cut-off to give forces convergence of less than 0.001 eV$\cdot$\r{A}$^{-1}$.  
Born effective charges and phonon calculations were performed within the density functional perturbation theory (DFPT) \cite{gonze1997} as implemented in \textsc{vasp}.
Spin-orbit coupling (SOC) was included to consider noncollinear magnetic configurations \cite{Hobbs2000}.
The polarization calculations were performed by using the Berry-phase approach, as implemented in the  \textsc{vasp} code \cite{VANDERBILT2000147}.
Phonon dispersions were post-processed in the \textsc{Phonopy} code \cite{phonopy}.
The atomic structure figures were elaborated with the \textsc{vesta} code \cite{vesta}.

%%%%%%%%%%%%%%%%
\begin{figure}[!t]
 \centering
 \includegraphics[width=7.0cm,keepaspectratio=true]{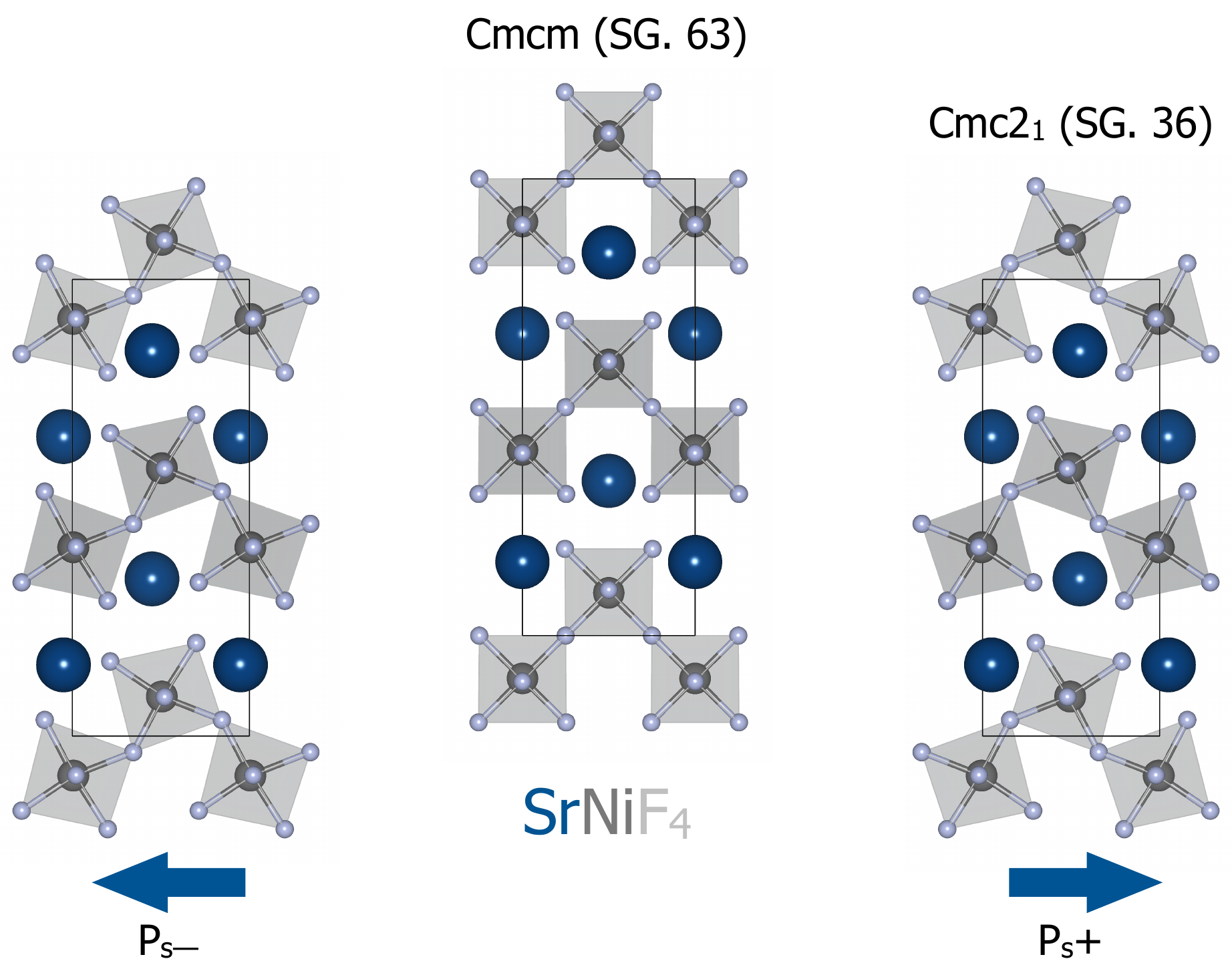}
 \caption{(Color online) Unit-cell structure of the high-symmetry $Cmcm$ paraelectric phase, as well as the low-symmetry polar $Cmc2_1$ space group, explained in terms of the polar $\Gamma_2^-$ mode that leads the group-to-subgroup transition. Moreover, here we show the positive and negative condensation of the modes in which, the reversal of the polarization is suggested.}
 \label{F1}
\end{figure} 
%%%%%%%%%%%%%%%%

%(a) Phonon-dispersion curve computed at the high-symmetry $Cmcm$ structure and (b) in the low-symmetry polar $Cmc2_1$. Here, the unstable modes are displayed as negative frequencies by notation. The latter phonon dispersions were computed along the high-symmetry points in the BZ equivalent in both phases. (c)

%%%%%%%%%%%%%%%%%%%%%%
\section{Results and discussion:}
\label{secIII}
\begin{figure}[!t]
 \centering
 \includegraphics[width=6.5cm,keepaspectratio=true]{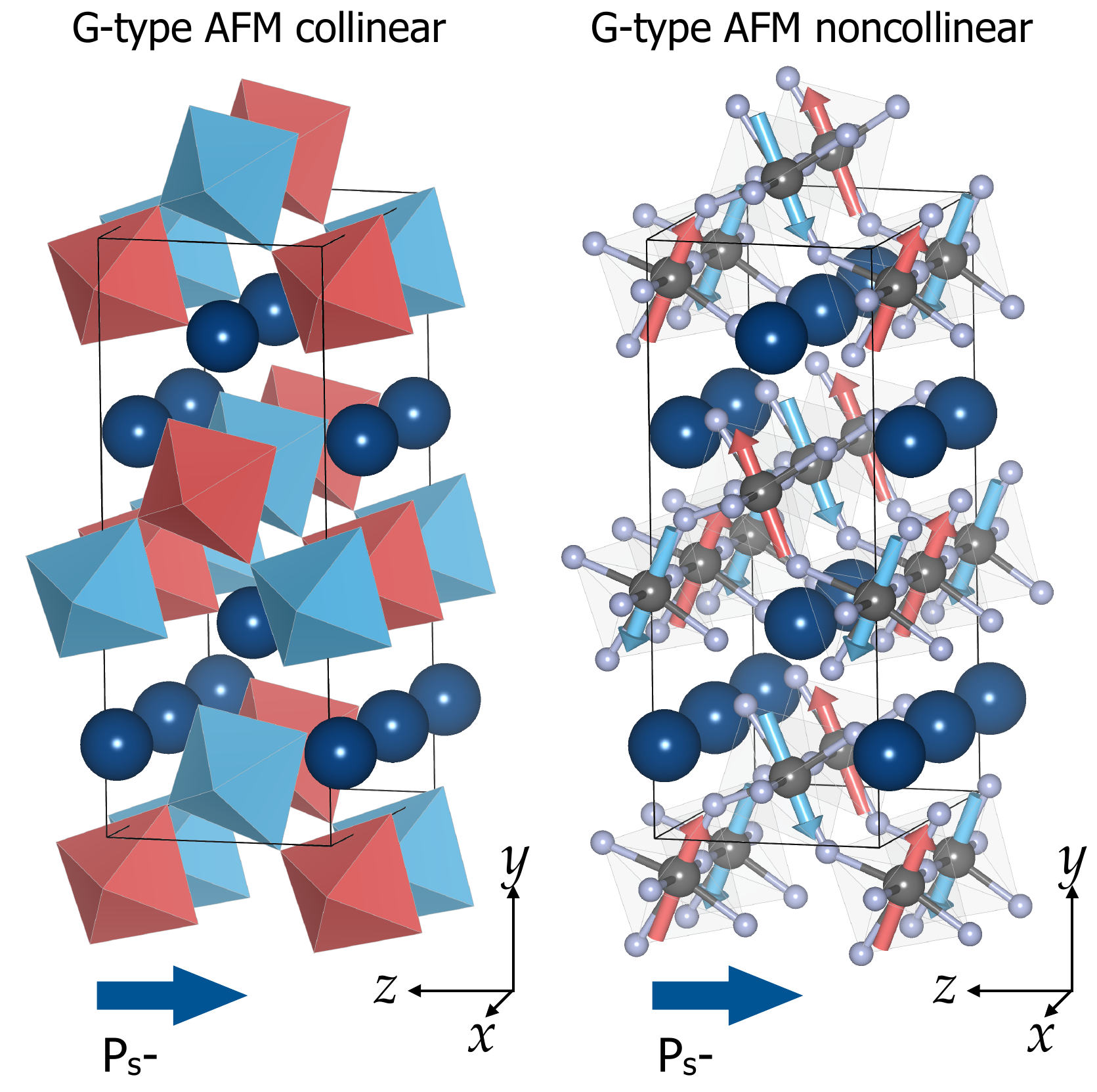}
 \caption{(Color online) $G$-type magnetic ordering in the collinear regime where the up- and down-magnetic moments orientation are shown as blue and red octahedra, respectively. The latter was observed for each Ni site within the NiF$_6$ octahedra. Here, we also show the noncollinear AFM magnetic ordering where the major magnetic moment, now represented by blue and red arrows, points toward the $y$-axis, and the weak AFM tilting is observed along the $z$-axis. The latter weak-AFM is coupled to the spontaneous ferroelectric polarization along the $z$-axis.}
 \label{F3}
\end{figure}

In the family of layered materials, SrNiF$_4$ follows the stoichiometric formulae $A_nB_nX_{3n+2}$ in which, $A$ = Sr, $B$ = Ni, $X$ = F, and $n$ = 2 as the other members of the Ba$M$F$_4$ family, as well as the LaTaO$_4$ \cite{LIU201631}.
This structure, observed in Fig. \ref{F1}, is characterized by the layered structure where the NiF$_6$ octahedral layers contain the Sr sites embedded into the free octahedral space.
Here, the Ni$^{+2}$ site holds a magnetically active 3$d^8$ occupation that, through the super-exchange interaction of the structure, leads to a $G$-type AFM magnetic ordering, see Fig. \ref{F3} discussed later in detail. 
Experimentally, the SrNiF$_4$ has been found to belong to the $Cmc2_1$ (SG. 36) polar space group. In this symmetry, the lattice parameters, measured by X-ray diffraction, are $a$ = 3.935 \r{A},  $b$ = 14.435 \r{A}, and  $c$ =   5.653 \r{A} \cite{SrNiF4-ref,BABEL198577}.
In our case, fully electronic and atomic relaxations lead us to lattice parameters of $a$ = 3.971 \r{A},  $b$ = 14.334 \r{A}, and  $c$ = 5.566 \r{A}. 
Here, the errors in the lattice parameters are below 1.5\% showing fair agreement with the experimental findings.

\begin{figure}[!b]
 \centering
 \includegraphics[width=8.8cm,keepaspectratio=true]{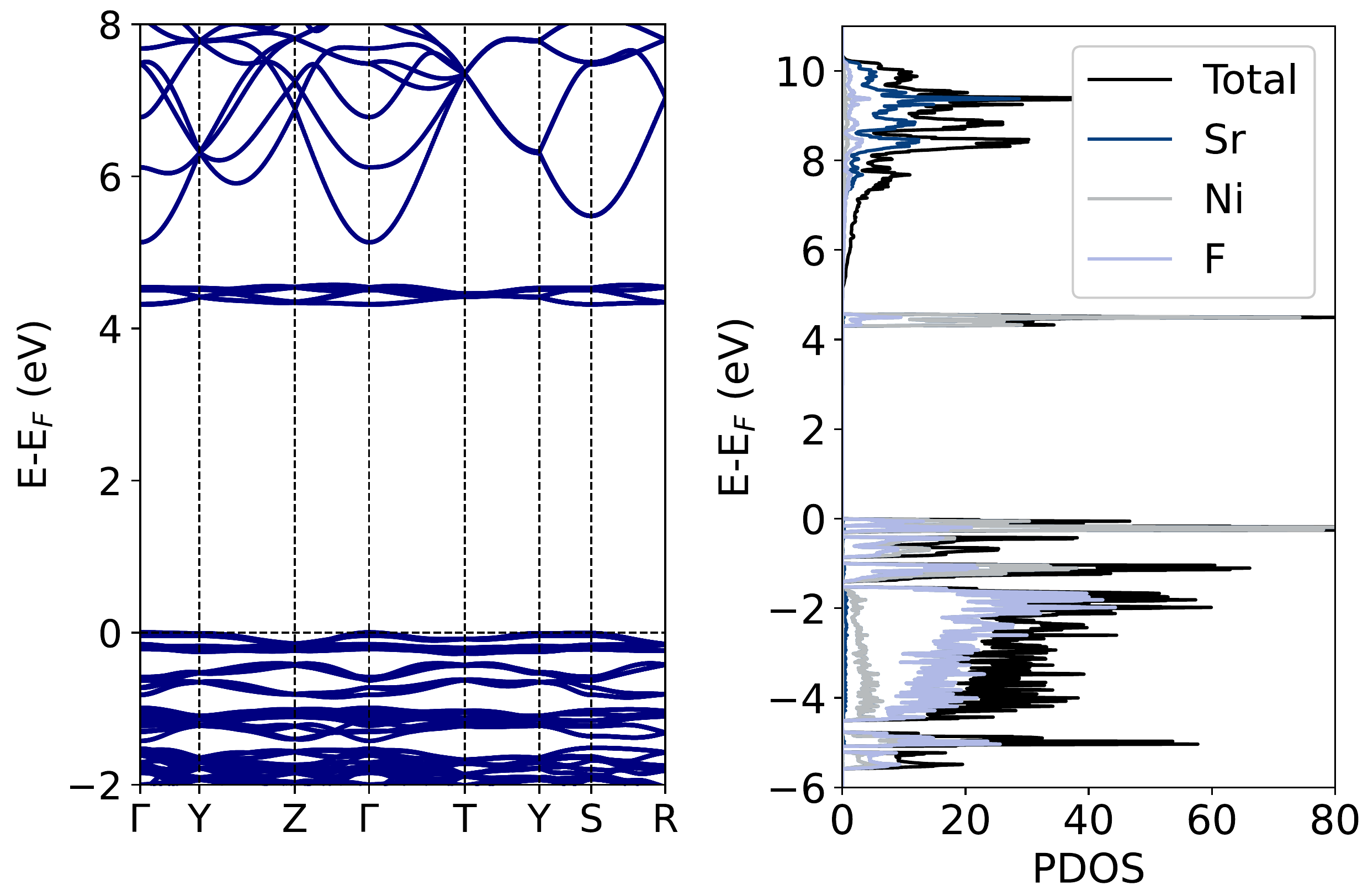}
 \caption{(Color online) Electronic band structure and atomically projected density-of-states, PDOS, obtained for the SrNiF$_4$ in the low-symmetry polar $Cmc2_1$ symmetry. The PDOS presented in state per eV.}
 \label{F1-1}
\end{figure}

\textcolor{black}{Aiming to provide more insights around the electronic structure of this layered fluoride, we have obtained the full electronic band structure, as well as the atomically projected density-of-states, both included in Fig. \ref{F1-1}. As it can be observed a wide bandgap energy characteristic of the highly ionic fluoride-like perovskite compounds.
Based on the projected-DOS, we can confirm that the Ni$+2$ leads to the 3$d$:$t_{2g}^6$$e_{g}^2$ electronic configuration in the SrNiF$4$ fluoride in which, an energy bandgap of 4.30 eV is observed. The latter in between the occupied Ni:3$d$($t_{2g}^6$$e_{g}^2$) and the unoccupied Ni:3$d$($e_{g}^2$) states. Moreover, such crystal-field splitting, characteristic of the NiF$_6$ octahedral coordination explains the observed magnetic moment per Ni atom. At an energy between 5 to 11 eV are observed the Sr states in agreement with the Sr($+2$) oxidation state. As such, the F-states are observed below the Fermi level, here at 0.0 eV by convention, as expected from the F($-1$) oxidation state in the SrNiF$_4$ fluoride.}

%%%%%%%%%%%%%%%%
%%%%%%%%%%%%%%%%

As reported before by some of the authors, the reduction of the $A$-site's atomic size, from Ba to Sr, leads to a considerable enhancement of the polarization and an increase of the $M$F$_6$ octahedral rotations \cite{Espitia_2020}. Therefore, in the SrNiF$_4$ case, it is highly expected to observe a polarization larger than in the BaNiF$_4$ counterpart, in which, $P_s$ = 6.8 $\mu$C$\cdot$cm$^{-1}$.
To be able to properly estimate the spontaneous polarization and explain the origin of the ferroelectric response, we obtained the vibrational frequencies at the BZ zone center. The latter is obtained in the hypothetical $Cmcm$ (SG. 63) high-symmetry structure which is the closest non-polar high-symmetry structure to be related with and connected by a group-to-subgroup transition as expected for the ferroelectric response.
In Table \ref{tab:modes} we present the obtained total Raman and IR modes. As it can be appreciated, the $\Gamma_2^-$ ($B_{1u}$) is unstable, here at negative frequencies by notation, and shows a frequency value of $\omega$ = $-$112.5 $i$cm$^{-1}$.
This unstable ($i.e.$ imaginary) mode suggests a symmetry-allowed transition from the $Cmcm$ high-symmetry structure based on this soft mode.
It is worth mentioning that the other vibrational modes are stable showing positive frequencies in the high-symmetry structure.
After extracting and analyzing the eigendisplacements associated with $\Gamma_2^-$ we observe that it involves the NiF$_6$ octahedral rotations, around the $x$-axis coupled with the Sr-sites polar displacements, along the $z$-axis.
As such, by condensation of the $\Gamma_2^-$ phonon mode, along the positive (negative) directions, it can be obtained the $P_s^+$ ($P_s^-$) polarization orientation and the symmetry is lowered up to the $Cmc2_1$ (SG. 36) non-centrosymmetric polar group.

%%%%%%%%%%%%%%%%
\begin{table}[!t]
\centering
\caption{DFT-PBEsol calculated BZ zone center phonon frequencies for the high-symmetry $Cmcm$ structure, as well as the polar $Cmc2_1$. As expected, in the $Cmcm$ the $\Gamma_2^-$ ($B_{1u}$) mode is unstable, here at negative frequency by notation, that leads the structure to the $Cmc2_1$.}
\begin{tabular}{c  c}
\hline
\hline
Mode  &   $Cmcm$, $\omega$ (cm$^{-1}$) \rule[-1ex]{0pt}{3.5ex}  \\
\hline
$B_{1u}$ &  $\bf{-113}$, 198, 230, 342, 529   \rule[-1ex]{0pt}{3.5ex}  \\
$B_{2u}$ &  74, 171, 242, 309, 432, 591 \rule[-1ex]{0pt}{3.5ex}  \\
$B_{3u}$ &  68, 176, 265, 407    \rule[-1ex]{0pt}{3.5ex}  \\
$A_{u}$ &  91, 248     \rule[-1ex]{0pt}{3.5ex}  \\
$B_{1g}$  & 96, 154, 315, 406   \rule[-1ex]{0pt}{3.5ex}  \\
$B_{2g}$ &  315     \rule[-1ex]{0pt}{3.5ex}  \\
$B_{3g}$ &  96, 122, 214, 278, 395    \rule[-1ex]{0pt}{3.5ex}  \\
$A_{g}$ &  101, 175, 254, 317, 459    \rule[-1ex]{0pt}{3.5ex}  \\
\hline
Mode &  $Cmc2_1$,  $\omega$ (cm$^{-1}$)  \rule[-1ex]{0pt}{3.5ex} \\
\hline
$B_{1}$ & 115, 199, 248 278, 467  \rule[-1ex]{0pt}{3.5ex}  \\ 
$B_{2}$ & 61, 108, 171, 178, 209, 242, 272, 337, 387, 425, 501 \rule[-1ex]{0pt}{3.5ex}  \\ 
$A_{1}$ & 89, 117, 145, 176, 201, 240, 267, 306,  348, 445, 458 \rule[-1ex]{0pt}{3.5ex}  \\ 
$A_{2}$ & 96, 153, 176, 222, 285, 267 \rule[-1ex]{0pt}{3.5ex} \\
\hline
\hline
\end{tabular}
\label{tab:modes}
\end{table}
%%%%%%%%%%%%%%%%

After full electronic and atomic relaxation within the DFT+$U$ - PBEsol approach, the total energy difference, between the nonpolar $Cmcm$ and the polar $Cmc2_1$ phase, is $\Delta E$= -209.5 meV$\cdot$f.u.$^{-1}$. The latter structure is obtained once the ($\Gamma_2^-$ + $\Gamma_1^+$) modes are fully condensed into the structure and this is allowed to relax. 
This finding is in agreement with the one observed in other compounds, such as BaZnF$_4$ in which, the $\Delta E$= -218.3 meV$\cdot$f.u.$^{-1}$ \cite{PhysRevB.93.064112}. Nonetheless, when only the $\Gamma_2^-$ is considered we observed an $\Delta E$= -120.3 meV$\cdot$f.u.$^{-1}$ suggesting the importance of the phonon coupling induced by the $\Gamma_1^+$ into the structure.
In Table \ref{tab:modes} we also include the computed frequencies of the fully electronically and atomically relaxed polar $Cmc2_1$ phase. As expected when compared to the experimental reports, the $Cmc2_1$ is fully stable at the BZ zone center frequencies.

\begin{figure}[!b]
 \centering
 \includegraphics[width=8.0cm,keepaspectratio=true]{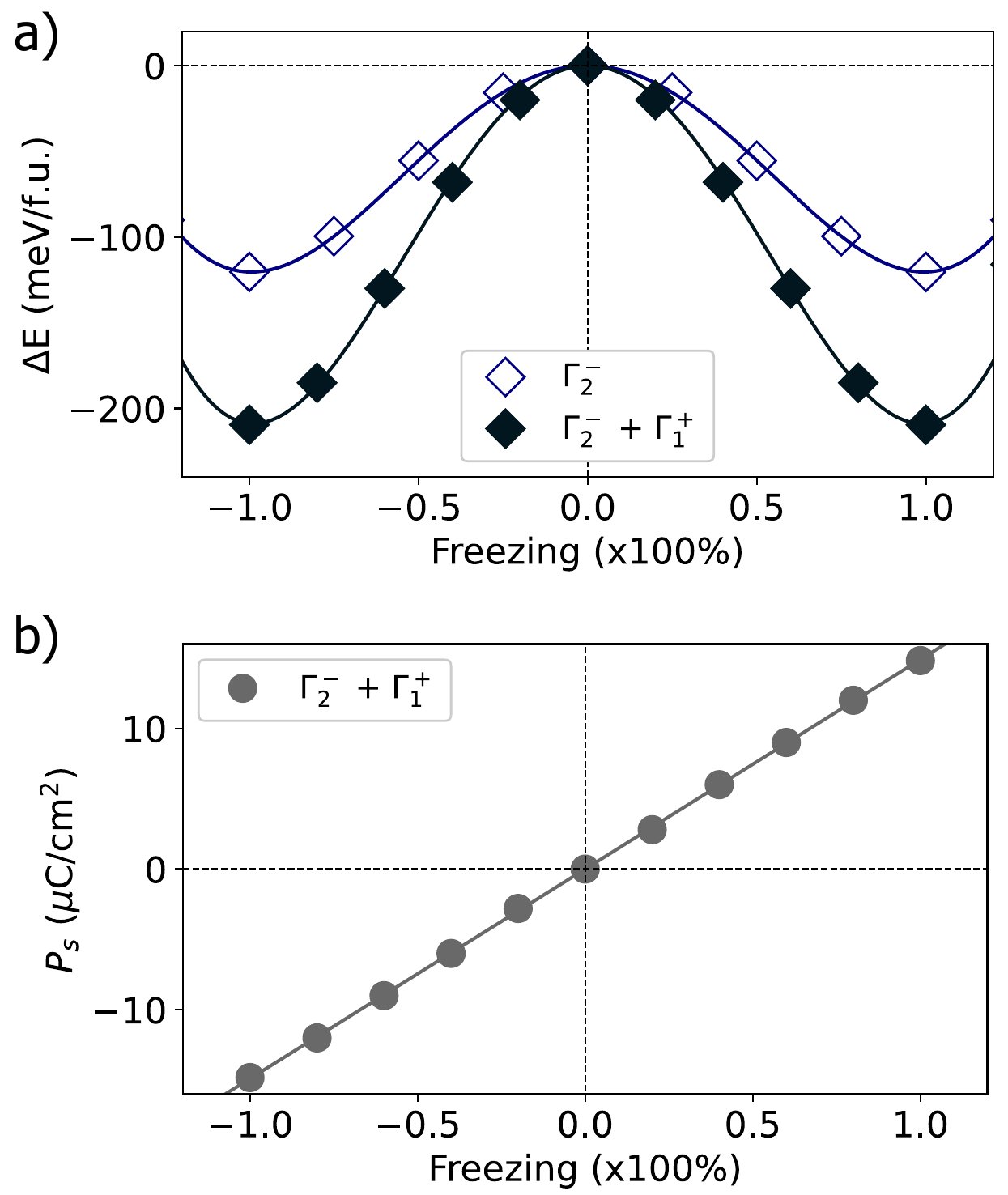}
 \caption{(Color online) (a) Double-well energy obtained after the condensation of the displacements associated with the $\Gamma_2^-$ and $\Gamma_2^-$ + $\Gamma_1^+$ phonon modes into the paraelectric high-symmetry $Cmcm$ structure. Thus, at the top of the well and 0\% displacements, the structure lies in the $Cmcm$ structure, whereas at the bottom of the energy wells we have the polar $Cmc2_1$ phases with equivalent positive and negative orientations of the polar displacements. (b) Spontaneous polarization computed as a function of the ($\Gamma_2^-$ + $\Gamma_1^+$) phonon modes condensation. At the 100\% freezing, the $Cmc2_1$ phase is obtained after full electronic and atomic relaxation.}
 \label{F2}
\end{figure}

In Fig. \ref{F2}a we present the potential energy profile obtained after the ($\Gamma_2^-$) mode condensation. As expected, a double-well energy profile is obtained confirming the group-to-subgroup transition allowed symmetry. Here, at the top of the energy profile, at 0\% of freezing, we have the $Cmcm$ high-symmetry non-polar phase, as soon as the eigendisplacements associated with the modes are frozen into the structure, the total energy is lowered up to a minimum of $-$120.3 meV$\cdot$f.u.$^{-1}$. At the bottom of the well, we have the $Cmc2_1$ low-symmetry polar phase.
All the latter explains the symmetry and lattice-dynamics allowed polar phase and potential switching path necessary in ferroelectric compounds.
Interestingly, if the structure is allowed to fully relax, the ($\Gamma_2^-$ + $\Gamma_1^+$) modes are observed in the lower structure. 
This is a consequence of the phonon-phonon coupling that brings the stable $\Gamma_1^+$ into the structure. This coupling reinforces and enhances the polar displacements and leads to a larger difference in energy, which in this case is $-$209.5 meV$\cdot$f.u.$^{-1}$, as it can be observed also in Fig. \ref{F2}(a).

%%%%%%%%%%%%%%%%
%%%%%%%%%%%%%%%%

Now, to account for the spontaneous polarization in the polar low-symmetry phase, we compute the $P_s$ by employing the Berry-phase approach \cite{VANDERBILT2000147}. The latter calculation was performed by tracing the evolution of the polar displacements as a function of the modes' condensation. 
In Fig. \ref{F2}b we present the computed spontaneous polarization values, in $\mu$C$\cdot$cm$^{-2}$, as a function of the eigendisplacements associated with the ($\Gamma_2^-$ + $\Gamma_1^+$) modes. 
As expected, we obtain values of positive (negative) saturation polarization for positive (negative) condensation of the eigendisplacements showing the possible ferroelectric switching path. 
Here, we obtained, in the fully relaxed $Cmc2_1$ phase, a polarization value of 14.8 $\mu$C$\cdot$cm$^{-2}$. 
This value is considerably larger than other family members, see for example Table \ref{tab:1} for comparison.

%%%%%%%%%%%%%%%%
\begin{table}[!b]
\caption{Values of spontaneous polarization and ionic atomic $M^{2+}$ radii size \cite{Shannon1976} for the reported polar and ferroelectric members of the Ba$M$F$_4$ family. In this cases, $R_{Ba^{2+}}$ = 149 pm, $R_{Sr^{2+}}$ = 132 pm and $R_{F^{1-}}$ = 119 pm. Here, it is observed a trend in the polarization as a function of the octahedral cation size. The latter is in agreement with the geometric proper ferroelectricity phenomenon found in these fluoride compounds. *In these compounds BaMnF$_4$ and BaFeF$_4$ no experimental ferroelectric switching has been achieved. In the last column, we present the total energy barrier values for all the compounds considering the $Cmcm$ high-symmetry reference.}
\begin{center}
\centering
\begin{tabular}{c c c c}
\hline
\hline
Compound &  $R_{M^{2+}}$  (pm)   & P$_s$ ($\mu$C$\cdot$cm$^{-2}$)   &  $\Delta$E (meV$\cdot$f.u.$^{-1}$)  \rule[-1ex]{0pt}{3.5ex} \\
\hline
SrNiF$_4$ &  83 &  14.8   &  210 \rule[-1ex]{0pt}{3.0ex} \\
BaNiF$_4$ &  83 &  6.8 \cite{PhysRevB.74.024102}   &  28 \cite{PhysRevB.74.024102}    \rule[-1ex]{0pt}{3.0ex} \\
BaMgF$_4$ &  86 &  9.9   &    133  \cite{PhysRevB.93.064112}   \rule[-1ex]{0pt}{3.0ex} \\
BaCuF$_4$ &  87 &  10.9 \cite{PhysRevLett.121.117601}  &  27 \cite{PhysRevLett.121.117601}   \rule[-1ex]{0pt}{3.0ex} \\
BaZnF$_4$ &  88 &  12.2   &  218   \cite{PhysRevB.93.064112}  \rule[-1ex]{0pt}{3.5ex} \\
BaCoF$_4$ &  88.5 &  9.0 \cite{PhysRevB.74.024102} &  58 \cite{PhysRevB.74.024102}   \rule[-1ex]{0pt}{3.0ex} \\
*BaFeF$_4$ &  92 &  10.9 \cite{PhysRevB.74.024102} &  122  \cite{PhysRevB.74.024102}   \rule[-1ex]{0pt}{3.5ex} \\
*BaMnF$_4$ &  97 &  13.6 \cite{PhysRevB.74.024102} &  191  \cite{PhysRevB.74.024102}   \rule[-1ex]{0pt}{3.0ex} \\
\hline
\hline
\end{tabular}
\end{center}
\label{tab:1}
\end{table}
%%%%%%%%%%%%%%%%

To compare the properties of the SrNiF$_4$ with the rest of the members of the family, we also include in Table \ref{tab:1} the ionic $M^{2+}$ radii, spontaneous $P_s$ polarization, and the computed ferroelectric switching well $\Delta E$ energy difference for several reported compounds. Here, we can observe that the energy barrier of the SrNiF$_4$ compound is close to the experimentally observed BaZnF$_4$, besides, the spontaneous polarization is the largest among the family, as commented before. 
It is important to note that, in the case of BaFeF$_4$ and BaMnF$_4$, the ferroelectric switching has not been experimentally observed and only piezoelectric and polar responses are suggested. The latter is possibly related to the $M_2^+$ atomic radius size that could be preventing the polarization reversal. Nevertheless, further studies are needed to unveil the nature of the neglected polarization switching.
We have also computed the Born effective charges, in Table \ref{tab:charges} at \ref{AppB}, to look for a possible dynamical charge transfer that can be related to the ferroelectric phenomenon. The $Z^*$ charges are computed as the derivatives of the polarization with respect to atomic displacements for each atomic symmetry site in the $Cmc2_1$ polar phase. 
As observed, the Born effective charges, $Z^*$, for the Sr, Ni, and F sites are close to their nominal +2, +2, and -1 $e^-$ charge values, respectively. These findings support the geometrically-driven ferroelectric response, as in the case of similar systems \cite{LIU201631, PhysRevB.89.104107, PhysRevB.74.024102, PhysRevB.74.020401,doi:10.1021/acsami.5b10814,PhysRevLett.121.117601,PhysRevLett.116.117202}.

%%%%%%%%%%%%%%%%
%%%%%%%%%%%%%%%%

Concerning the magnetic structure, \textcolor{black}{and after considering the FM, $A$-AFM, $C$-AFM, and $G$-AFM orderings in the SrNiF$_4$}, we found that the magnetic moment per Ni-site is $m$ = 1.755 $\mu_B$ per Ni site and the structure holds $G$-type AFM order. \textcolor{black}{The latter after obtaining energy differences, per formula unit, of $\Delta$E = -8,5 meV$\cdot$f.u.$^{-1}$, $\Delta$E = -33.4 meV$\cdot$f.u.$^{-1}$, and $\Delta$E = -40.7 meV$\cdot$f.u.$^{-1}$ for the $A$-AFM, $C$-AFM, and $G$-AFM magnetic orderings, respectively. All of these by considering the ferromagnetic state energy as a reference.}
Here, all the nearest neighbors of the Ni-sites order antiferromagnetically, as shown schematically for the collinear $G$-type AFM ordering presented in Fig. \ref{F3}. 
This is explained by the Ni--F--Ni magnetic exchange paths that favor the super-exchange interaction.
As in the BaNiF$_4$ case and other members of the family, in the SrNiF$_4$ the noncollinear ordering is allowed by symmetry and expected to be enhanced by the octahedral rotations. 
The octahedral rotation angle in SrNiF$_4$ is 20.7$^o$ compared to 14.0$^o$ in the BaNiF$_4$ counterpart.
After considering the noncollinear ordering into the calculations, as shown in Fig. \ref{F3}, and performing full electronic and atomic relaxation, we found that the main magnetic moment is $m_y$ = 1.752 $\mu_B$ per Ni atom and it is pointing towards the $b$-axis. 
Meanwhile, the weak-AFM moment is $m_z$ = 0.073 $\mu_B$ per Ni site and is aligned along the $c$-axis. 
Interestingly, the latter weak AFM moment is smaller than our computed component in the BaNiF$_4$ of $m_z$ = 0.102 $\mu_B$.
This finding is contradictory based on the enhanced rotations, in the SrNiF$_4$, induced by the smaller atomic radii size of Sr when compared to Ba. 
Although we observed a reduction in the expected canted component, the magnetoelectric character of the SrNiF$_4$ is confirmed. 
It is also worth mentioning that the reported Ne\`{e}l temperature of the SrNiF$_4$ lies in $T_N$ = 100 K \cite{SNF-1984} suggesting a multiferroic/magnetoelectric response below such temperature. 
Despite these low temperatures, the ferroelectric behavior is expected to appear well above room temperature and the large coupling between the magnetic and polar degrees of freedom through the $\Gamma_2^-$ phonon mode makes these compounds suitable cases of study for 2D-like multiferroic/magnetoelectric phenomenon.

\textcolor{black}{Finally, we have explored the potential effect of the Coulomb, $+U$, correction in the structural, electronic, and magnetic properties of the polar $Cmc2_1$ structure, see \ref{AppC}.
We observed an expected opening of the bandgap energy when the $+U$ value is increased. As such, an slight increase of the magnetic moment per Ni atom is also appreciated. The latter is induced by the rise in the Ni:3$d$ states localization as a function of the $+U$ value. 
On the other hand, the lattice parameters remains almost constant as a function of the $+U$ value suggesting an weak influence of such parameter into the structural degrees of freedom.}

\section{Conclusions and general remarks:}
\label{conclusions}
We theoretically investigated, by means of first-principles calculations in the framework of Density-funcional theory, the structural, ferroelectric, and magnetic properties of the SrNiF$_4$ fluoride layered compound. 
The latter compound stands as an unexplored member of the $AM$F$_4$ ($A$ = Ba and Sr, $M$ = Fe, Co, Mn, and Ni) of multiferroic/magnetoelectric layered fluorides family.
We found that, due to the geometric nature of the polarization response, the spontaneous polarization has a value as large as $P_s$ = 14.8 $\mu$C$\cdot$cm$^{-2}$. 
This polarization is one of the largest in the family and is explained in terms of the large degree of rotations and small Sr$^{2+}$ cation's size, in SrNiF$_4$, when compared with the BaNiF$_4$ compound.
Concerning the magnetic response, we observed a $G$-type antiferromagnetic response in the collinear regime explained in terms of the Ni--F--Ni superexchange interaction paths for the interlayer structure.
Based on the symmetry considerations, the non-collinear magnetic structure is also allowed and we found a main antiferromagnetic ordering along the $y$-axis with a weak AFM ordering along the $z$-axis entangled with the spontaneous polarization axis.

%%%%%%%%%%%%%%%%%%%%%%
\section*{Acknowledgments:}
\label{acknowledgements}
The calculations presented in this paper were carried out using the Grid UIS-2 experimental testbed, being developed under the Universidad Industrial de Santander (SC3-UIS) High Performance and Scientific Computing Centre, development action with support from UIS Vicerrector\'ia de Investigaci\'on y Extensi\'on (VIE-UIS) and several UIS research groups as well as other funding resources.
A. C. Garcia-Castro and W. Ibarra-Hernandez acknowledge grant No. 202303059C entitled Optimizaci\'on de las Propiedades Termoel\'ectricas Mediante Tensi\'on Biaxial en la Familia de Materiales Bi$_4$O$_4$Se$X_2$ ($X$ = Cl, Br, I) Desde Primeros Principios supported by the LNS - BUAP. 
The authors thankfully acknowledge computer resources, technical advice, and support provided by Laboratorio Nacional de Superc\'omputo del Sureste de M\'exico (LNS), a member of the CONAHCYT national laboratories, with project No. 202303059C.

\section*{CRediT authorship contribution statement}
\textbf{W. Ibarra-Hernández} and \textbf{A. Bautista-Hernández}: Calculations and analysis of the data. \textbf{A. C. Garcia-Castro}: Supervision, calculations, analysis, and writing of the original draft.

\section*{Authors' contributions}
All of the authors were involved in the preparation and development of the manuscript.
Moreover, all of the authors read and approved the final manuscript.

\section*{Data availability}
The data that support the findings of this study are available from the corresponding author upon reasonable request.

\section*{Conflict of interest}
There are no conflicts to declare.

%%%%%%%%%%%%%%%%%%%%%%%%%%%%%%%%%%%%%%%%
\appendix

%\section{Phonon dispersion curves computed in the high- ($Cmcm$) and low-symmetry ($Cmc2_1$):}\label{AppA}
%In the following, we present the phonon-dispersion curves obtained for $Cmcm$ and the $Cmc2_1$ structure, see Fig. \ref{F5}

%%%%%%%%%%%%%%%%
%\begin{figure}[!h]
% \centering
% \includegraphics[width=6.0cm,keepaspectratio=true]{./F3.pdf}
% \caption{(Color online) Phonon-dispersion curves computed for the high- and low-symmetry $Cmcm$ and $Cmc2_1$, respectively.}
% \label{F5}
%\end{figure}
%%%%%%%%%%%%%%%%

\section{Born effective charges:}\label{AppB}
Aiming to support the origin of the polar behavior in the SrNiF$_4$, we computed Born effective charges, $Z^*$, for the Sr, Ni, and F  sites, in the $Cmc2_1$ polar phase, see Table \ref{tab:charges}. As it can be appreciated, $Z^*$ are close to their nominal value in each case suggesting a major ionic bonding phenomenon and a geometric-like polar distortion.

\begin{table}[!h]
\centering
\caption{Computed Born effective charges, $Z^*$ in $e^-$ units, for the Sr, Ni, and F  sites and obtained in the $Cmc2_1$ polar phase.}
\begin{tabular}{c  c  c | c  c  c}
\hline
\hline
 &  $Z^*_{Sr}$ & &  & $Z^*_{Ni}$  &   \rule[-1ex]{0pt}{3.5ex} \\
\hline
+2.671 & 0.000 & +0.000  & +2.117 &  0.000 & 0.000  \rule[-1ex]{0pt}{3.5ex}\\
 0.000 & +2.317 & -0.210  & 0.000 & +1.974 & -0.002   \rule[-1ex]{0pt}{3.5ex}\\
 0.000 & -0.167 &  +2.522  & 0.000 & -0.067 & +2.070  \rule[-1ex]{0pt}{3.5ex}\\
\hline
 &  $Z^*_{F1}$ & &  & $Z^*_{F2}$  &   \rule[-1ex]{0pt}{3.5ex} \\
\hline
-1.056 &  0.000 &  0.000  & -1.115 &  0.000 &  0.000  \rule[-1ex]{0pt}{3.5ex}\\
 0.000 & -1.341 & +0.033  &  0.000 & -0.980 & +0.170   \rule[-1ex]{0pt}{3.5ex}\\
 0.000 & +0.027 & -1.121  &  0.000 & +0.167 & -1.355  \rule[-1ex]{0pt}{3.5ex}\\
 \hline
  &  $Z^*_{F3}$ & &  & $Z^*_{F4}$  &   \rule[-1ex]{0pt}{3.5ex} \\
 \hline
-0.854 &  0.000 &  0.000  & -1.763 &  0.000 &  0.000  \rule[-1ex]{0pt}{3.5ex}\\
 0.000 & -1.246 & -0.335  &  0.000 & -0.723 & -0.083   \rule[-1ex]{0pt}{3.5ex}\\
 0.000 & -0.360 & -1.202  &  0.000 & -0.089 & -0.913  \rule[-1ex]{0pt}{3.5ex}\\
\hline
\hline
\end{tabular}
\label{tab:charges}
\end{table}

\section{Effect of the Coulomb, $+U$, correction:}\label{AppC}
In magnetically active compounds, and strongly correlated crystals, the Coulomb $+U$ exchange can influence the structural, electronic, and magnetic properties through the spin-lattice coupling. As such, we have obtained estimated the effect of the Coulomb $+U$ parameter by computing the lattice parameters, the magnetic moment per Ni atom, and bandgap energy as a function of $U$, see Table \ref{CoulombU}.

\begin{table}[h]
\centering
\caption{$a$, $b$, and $c$ lattice parameter, magnetic moment per nickel site, and bandgap energy computed as a function of the $+U$ Coulomb parameter in the polar $Cmc2_1$ phase.}
\begin{tabular}{c c c c | c | c}
\hline
\hline
$U$ (eV) & $a$ (\r{A}) & $b$ (\r{A}) & $c$ (\r{A}) & $m$ ($\mu_B$ $\cdot$ Ni$^{-1}$) & $E_g$ (eV)  \rule[-1ex]{0pt}{3.5ex}\\ 
\hline
0.0 & 3.973 & 14.245 & 5.540 & 1.582 & 1.61 \rule[-1ex]{0pt}{3.5ex}\\
1.0 & 3.971 & 14.266 & 5.545 & 1.643 & 2.18 \rule[-1ex]{0pt}{3.5ex}\\
2.0 & 3.970 & 14.278 & 5.548 & 1.688 & 2.87 \rule[-1ex]{0pt}{3.5ex}\\
3.0 & 3.969 & 14.286 & 5.552 & 1.724 & 3.66 \rule[-1ex]{0pt}{3.5ex}\\
4.0 & 3.969 & 14.288 & 5.554 & 1.755 & 4.30 \rule[-1ex]{0pt}{3.5ex}\\
5.0 & 3.968 & 14.298 & 5.556 & 1.783 & 4.97 \rule[-1ex]{0pt}{3.5ex}\\
6.0 & 3.966 & 14.299 & 5.557 & 1.808 & 5.57 \rule[-1ex]{0pt}{3.5ex}\\
\hline
\hline
\end{tabular}
\label{CoulombU}
\end{table}

%\section*{References}

\bibliography{library}

\end{document}